\documentclass[12pt]{iopart}

\usepackage{bm}
\usepackage{iopams}
\usepackage{setstack}
\usepackage{graphicx}
\usepackage{amsthm}

\begin{document}
\title{Spin-orbit interaction in three-dimensionally bounded semiconductor nanostructures}
\author{Eduard Takhtamirov} 
\address{M$^2$NeT Laboratory, Wilfrid Laurier University, 75 University Avenue West, Waterloo, ON, N2L 3C5, Canada}
\ead{ed.takhtamirov@gmail.com}
\author{Roderick V. N. Melnik}
\address{M$^2$NeT Laboratory, Wilfrid Laurier University, 75 University Avenue West, Waterloo, ON, N2L 3C5, Canada, and}\address{BCAM, Bizkaia Technology Park, 48160 Derio, Spain}
\ead{rmelnik@wlu.ca}

\begin{abstract}
The structural inversion asymmetry-induced spin-orbit interaction of conduction band electrons in zinc-blende and wurtzite semiconductor structures is analysed allowing for a three-dimensional (3D) character of the external electric field and variation of the chemical composition. The interaction, taking into account all remote bands perturbatively, is presented with two contributions: a heterointerface term and a term caused by the external electric field. They have generally comparable strength and can be written in a unified manner only for 2D systems, where they can partially cancel each other. For quantum wires and dots composed of wurtzite semiconductors new terms appear, absent in zinc-blende structures, which acquire the standard Rashba form in 2D systems.
\end{abstract}

\maketitle

\section{Introduction}

Systems with lowered space symmetry are generally characterized by spin-split energy states. This is a manifestation of the relativistic interaction of moving magnetic angular momentum with electric field \cite{Landavshitz}. In the physics of semiconductor nanostructures, we conventionally identify two reasons for the effect: lack of inversion symmetry of the unit cells of the constituent materials \cite{Dresselhaus,Casella,Rashba1,Rashba2} and the presence of the structural inversion asymmetry (SIA) on macroscopic scale, much larger than the unit cell \cite{Bychkov,Silva}.

Major results in this field have been obtained by using only symmetry arguments, with the method of invariants \cite{Luttinger}, which has become the most efficient instrument to studying electron states in semiconductors \cite{Bir-Pikus}. However, this method has a drawback, for its phenomenological nature does not yield a distinct link between bulk materials and heterostructures. Each reduction in dimensions, and more generally, each loss of elements of symmetry including a result of the application of strong enough external fields, creates a new system that calls for an independent analysis \cite{Dandrea,Bernardini,Zunger}.

Another successful phenomenological tool, the $\mathbf {k \cdot p}$ method, the only extra requirement of which is mean-field approximation \cite{Bir-Pikus}, does provide the missing link along with its limits of applicability \cite{Takhtamirov}. For two-dimensional (2D) electron systems, this method has already proved to be capable of specifying the SIA mechanisms of the spin-orbit interaction and identifying them as the heterointerface induced and external electric field induced \cite{Silva,Leibler-so,Vasko,Gerchikov,Pfeffer}. Knowledge of their strengths is crucial to band structure engineering aimed at manipulation with the spin degree of freedom, which finds important applications \cite{Loss,Gurung, Benjamin}.

An interesting situation holds for quantum dots. Their lateral confinement is usually modelled as parabolic, validating the inclusion of only 1D interface spin-orbit interaction, see, e.g., \cite{Sanjay}. However, attempts to make allowance also for the spin-orbit interaction due to a lateral field have been made in the past \cite{Moroz}. This may be decisive for quantum dots based on piezoelectric materials such as GaN/AlN, where complex distributions of strain generate 3D pictures of strong (of the order of $10^6$~V/cm) `external' electric fields \cite{Andreev,Pan,Barettin,Patil1,Patil2}. In addition to structures with 3D confinement, even nominally 1D and 2D electron systems have their realizations in arrangements with boundaries and contacts inducing electric fields acting on the interior of the systems. That is why a 1D spin-orbit interaction model may fail there. Indeed, it may underestimate by an order of magnitude \cite{Litvinov} the electron spin splitting in GaN/AlN quantum wells, see e.g. \cite{Lisesivdin}.

We use the $\mathbf {k \cdot p}$ method to systematically include the effect of an external electric field of arbitrary profile and general 3D variation in chemical composition on the spin-orbit interaction in zinc-blende and wurtzite heterostructures. Previous studies dealt with either 3D composition profile and arbitrary external electric field, considering them in detail only for cubic materials \cite{Leibler-so}, or with 1D electric fields for wurtzite materials \cite{Litvinov}. In the former work, the basis functions comprised spin-orbit interaction. As a consequence, two different effects, describing the position-dependent effective mass and governing the spin-orbit interaction, combined, and some of the spin-orbit terms were not singled out. An incomplete picture of the spin-orbit interaction resulted in the conclusion that the heterointerface contribution is small if the spin-orbit interaction parameter weakly varies with composition. We find it more convenient to use the original effective-mass method's spinless basis \cite{Luttinger-Kohn} to obtain all the necessary terms. After analysis of them, we found that the heterointerface contribution should always be taken into account.

The study of wurtzite \cite{Litvinov}, borrowing the method used for zinc-blende materials \cite{Silva,Gerchikov,Pfeffer}, was limited to the eight-band Kane-type model \cite{Kane}. In that method, small (valence-band) envelope functions are excluded, resulting in a conduction band Hamiltonian that parametrically depends on its own eigenenergy \cite{Suris}. Apart from neglecting remote bands whose contributions have never been evaluated (with questionable relevance for wide-bandgap materials), such a Hamiltonian has applicability problems, e.g. when time-dependent external fields are considered. Our resulting SIA spin-orbit interaction terms for conduction band states near the Brillouin zone centre in zinc-blende and wurtzite semiconductors are Hermitian and energy-independent. Any remote bands can be taken into account if proper material parameters are known.

This paper is organised as follows. In section~2, we make a perspective analysis of the heterointerface- and external electric field-induced SIA mechanisms and find that they should generally be considered on equal footing. In section~3, we introduce a multi-band system of en\-velope-function equations used in section~4 to derive spin-orbit terms entering conduction band envelope-function equations, with details being given in appendix~A. The general expressions for the spin-orbit interaction terms are then analysed for zinc-blende and wurtzite heterostructures. We discuss the results in section~5.

\section{Heterointerface- and external electric field-induced terms: a comparison}

The heterointerface spin-orbit contribution first appears in the third order \cite{Leibler-so} of the L\"owdin perturbation scheme \cite{Bir-Pikus,Lowdin}. It is proportional to the difference in the spin-orbit interaction parameters for the semiconductors of the heterojunction. Analysing the `exact' expression for the parameter of the interface term in the Kane model \cite{Silva,Gerchikov,Pfeffer}, we see that there should also be a contribution due to the valence band offset, present even when the spin-orbit interaction parameter does not vary with composition. Such a term, proportional to the band offset and the spin-orbit interaction parameter, will be available only if we consider the fourth order of the perturbation scheme.

The electric field-induced contribution, which is proportional to the external electric field and the spin-orbit interaction parameter, arises only in the fourth order of the perturbation scheme if the Kohn-Luttinger basis functions \cite{Luttinger-Kohn} do not include spin-orbit interaction. Otherwise, Leibler showed that inclusion of the spin-orbit interaction in the basis makes both contributions present as third-order corrections \cite{Leibler-so}. For typical heterostructures, the spin-orbit interaction energy is less or of the order of the band offsets. This, together with conventional arguments discussed in section~4, makes the original method's spinless basis \cite{Luttinger-Kohn} preferable. Then, such a perturbative classification of the interface and electric field-induced terms might indicate that the former term is stronger than the latter.

To find out whether this is so, let us consider them for a quantum well grown along the $z$-axis and having for simplicity only one heterojunction located at $z=0$ (see figure 1). With only Rashba-type spin-orbit contributions present, the effective mass equation for the electron envelope functions $\Phi_n$, which correspond to the steady state eigenenergies $\epsilon_n$, can be written as $H_{2D}\Phi_n = \epsilon_n \Phi_n$, with the Hamiltonian:
\begin{equation}\label{H2D}
H_{2D} = \frac{\hbar^2{\bf k}^2}{2m^*}+U(z)+W(z)+ \left( R_1 \delta \left(z\right) + R_2 \frac {d W(z)}{dz} \right) \left( k_x \sigma_y - k_y \sigma_x \right),
\end{equation}
where $\hbar{\mathbf k}$ is the momentum operator, $m^*$ is the effective mass, $U(z)= \Theta (-z) \delta U_s$ and $W(z)$ are the potential energy of on electron in the crystalline potential and external scalar potential, respectively, $\Theta (z)$ is the Heaviside step function, $\delta U_s$ is the conduction band offset, $\delta(z)$ is the Dirac delta function, ${\boldsymbol\sigma} = (\sigma_x, \sigma_y,\sigma_z)$ is the Pauli matrix vector and $R_1$ and $R_2$ are material parameters. The term proportional to $R_1$ ($R_1$-term) defines the interface spin-orbit interaction. It is small by the parameter presented with a sum $\delta \Delta / E_g + \Delta \delta U_v / E^2_g$, where $\delta \Delta$ is the difference of the valence band spin-orbit splittings for the two materials forming the heterojunction, $\Delta$ is the valence band spin-orbit splitting for the quantum well material, $\delta U_v$ is the valence band offset and $E_g$ is the bandgap. The term proportional to $R_2$ ($R_2$-term) is the electric field-induced spin-orbit interaction. It is small by a parameter that is of the order of $\Delta \langle W\rangle / E^2_g$, where $\langle W\rangle$ is a characteristic potential energy of an electron in the external scalar potential.
\begin{figure}[t]
\includegraphics [width=8cm]{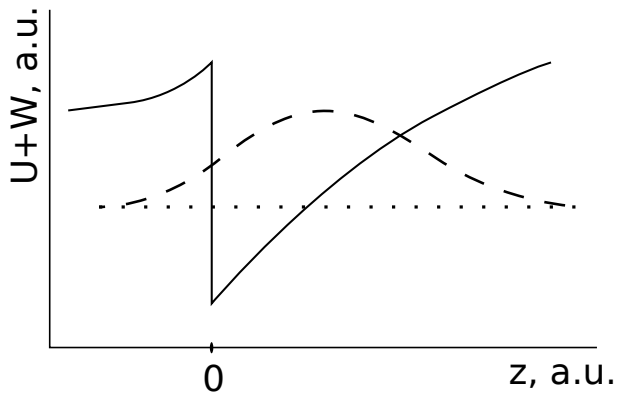}
\caption{A single-heterojunction potential well, showing the ground eigenenergy and its envelope function.
}
\label{f1}
\end{figure}

Following \cite{Pfeffer}, we consider the Hamiltonian $H_{2D0}$ of the zeroth-order approximation in the spin-orbit interaction:
\begin{equation}
H_{2D0} = \frac{\hbar^2{\bf k}^2}{2m^*}+U(z)+W(z).
\end{equation}
The corresponding envelope functions are designated as $\Phi_{0n} \equiv \vert n )$: $H_{2D0} \Phi_{0n} = \epsilon_n \Phi_{0n}$. Using the eigenfunctions $ \vert n )$ of the Hamiltonian $H_{2D0}$ and detailing diagonal matrix elements of the commutator $[k_z , H_{2D0}]$, which are zero, we immediately obtain
\begin{equation}\label{delta-E}
\delta U_s ( n \vert \delta \left(z\right) \vert n ) = ( n \vert \frac {d W(z)}{dz} \vert n ),
\end{equation}
holding for any subband $n$. Comparing this identity with the spin-orbit interaction part of (\ref{H2D}), we see that the terms proportional to $R_1$ and $R_2$ can both be written as a single interface contribution, found by Pfeffer and Zawadzki \cite{Pfeffer}. We argue that the identity (\ref{delta-E}) suggests that both spin-orbit terms can also be written as a single one proportional to the external electric field. This makes them mutually indistinguishable if phenomena involving intersubband transitions are not of interest, so that only a single effective material parameter can be extracted from experiment. In our opinion, the form requiring knowledge of the average electric field is preferable. The electric field can easily be estimated from the electrostatics of the semiconductor system involved, but it is impossible to make a direct evaluation of the magnitude of the envelope function at the heterointerface without numerical simulations. Analysis of a two-heterointerface quantum well does not alter the above conclusion.

The identity (\ref{delta-E}) also helps us to make a simple comparison of the $R_1$- and $R_2$-terms, which involves the analysis of two small indicating parameters: $\delta \Delta / E_g + \Delta \delta U_v / E^2_g$ and $\Delta \delta U_s / E^2_g$, respectively. As typically $\vert \delta U_s \vert \sim \vert \delta U_v \vert$, we immediately conclude that if the spin-orbit interaction energies for the materials of the structure differ essentially, that is $\vert \delta \Delta\vert \sim \Delta$, the heterointerface term dominates, as found by Leibler \cite{Leibler-so}. Otherwise, if $\vert \delta \Delta \vert \ll \Delta$, both terms should generally be retained, contrary to the conclusion in \cite{Leibler-so}. As examples, consider two popular semiconductor pairs forming heterostructures: GaAs/AlAs and GaN/AlN, taking the material pa\-ra\-me\-ters from \cite{Vurgaftman}. For GaAs/AlAs, $\Delta \approx 0.34$~eV, $\delta \Delta \approx - 60$~meV, $E_g \approx 1.52$~eV, $\delta U_s \approx 1$~eV and $\delta U_v \approx -0.53$~eV. For the pair GaN/AlN, $\Delta \approx 14$~meV, $\delta \Delta \approx 5$~meV, $E_g \approx 3.5$~eV, $\delta U_s \approx 2$~eV and $\delta U_v \approx -0.7$~eV. The indicating parameters are comparable, $\vert \delta \Delta \vert \ll \Delta$, so we should keep both terms. Below, in section~4, we produce a more accurate evaluation for these two semiconductor pairs and show that both systems have an `accidental' set of parameters leading to somewhat different conclusions. In the above estimates, we do not take into account remote bands whose effect has never been evaluated. For electron states in quantum wires and dots, the $R_1$- and $R_2$-type terms cannot be reduced to a unified form because the operators $k_x$ and/or $k_y$ do not commute with the Hamiltonian, and the identity (\ref{delta-E}) cannot be directly used in the spin-orbit interaction terms of the Hamiltonian.

In summary, the heterointerface- and external electric field-induced terms can be of comparable strength and, while being mutually indistinguishable for 2D electron systems, they cannot be written in a unified form for quantum wires and quantum dots. These make them be discrete and equally important.

\section{The multi-band system of envelope-function equations}

We now proceed to derive the SIA-governed spin-orbit terms using the $\mathbf {k \cdot p}$ method. We consider a heterostructure composed of two semiconductors with the net single-particle potential energy of electron, which will be called `potential' for brevity, $U = U\left( {\bf r}\right)$:
\begin{equation}
U=U_1+f \left[ U_2-U_1\right] \equiv U_1+f \delta U,
\end{equation}
where $U_1 = U_1\left( {\bf r}\right) $ and $U_2 = U_2\left( {\bf r}\right)$ are the periodic lattice potentials of the nominally potential well material and the barrier material, respectively. We suppose that the phenomenological function $f \equiv f\left( {\bf r}\right)$, which defines the profile of the structure, is of the order of unity or less even at the heterointerfaces \cite{Leibler-so,Leibler}. Ideally, $f$ can even be a step-like function taking the values $f=0$ in the region of the potential well material, and $f=1$ in the region of the barrier material \cite{Takhtamirov}. It can have a variation in 1D, 2D or 3D to represent a potential well, a quantum wire or a quantum dot, respectively. The final expression will have a local character allowing plain generalization for an arbitrary composition. We set the only requirement that the semiconductors composing the structure are not too dissimilar, so that $\delta U$ can be treated as a small perturbation as compared to the basis potential $U_1$.

In the mean-field approximation, the Schr\"odinger equation with the relativistic spin-orbit interaction term is \cite{Landavshitz}:
\begin{equation}
\left( \frac{\hbar^2{\bf k}^2}{2m_0}+U+\frac{\hbar^2 \left[ \boldsymbol\nabla U \times {\mathbf k} \right] \cdot \boldsymbol\sigma} {4m_0^2c^2} +W \right) \Psi \left( {\bf r}\right)
=\epsilon \Psi \left( {\bf r}\right) .
\end{equation}
Here $m_0$ is the free electron mass and $c$ is the velocity of light in vacuum. The external scalar potential $W = W\left( {\bf r}\right)$ is weak, and we neglect its direct relativistic effect. 

Dealing with states near the Brillouin zone centre, it is convenient to use the complete set of Kohn-Luttinger functions \cite{Luttinger-Kohn}, with the unit-cell normalized periodic parts $u_{n0} = u_{n0}\left( {\bf r}\right) \equiv \vert n \rangle$ specified as
\begin{equation}
\left( \frac{\hbar^2{\mathbf k}^2}{2m_0}+U_1 \right) \vert n \rangle =\epsilon _{n0} \vert n \rangle, \label{un0}
\end{equation}
where $\epsilon _{n0}$ is the $n$th band edge energy. The relativistic effect of the potential $U_1$ is not included in the basis, but processed as a perturbation.
We define the $n$th band envelope function in $\mathbf r$-representation as $A_n = A_n\left( {\bf r}\right)$ with
\begin{equation}\label{EF}
\Psi \left( {\bf r}\right) =\sum_nA_n \, \vert n \rangle,
\end{equation}
where the summation is over all bands.

Consider the following ${\mathbf k\cdot \mathbf p}$ system in $\mathbf r$-representation, which is obtained by treating the functions $f\left( {\mathbf r}\right)$ and $W\left( {\mathbf r}\right)$ as `gentle' \cite{Luttinger-Kohn}. This is a standard procedure of the envelope-function method \cite{Leibler}, which neglects all `central cell'-like corrections due to a rapid variation in the function $f\left( {\mathbf r}\right)$ at the heterointerfaces \cite{Takhtamirov}.
\begin{eqnarray}\label{kp}
\eqalign{
\left( \epsilon _{n0} 
+ \frac{\hbar ^2 {\mathbf k}^2}{2m_0} + W\left( {\mathbf r} \right)
\right) A_n + \sum_{n^{\prime }} \frac{ \hbar {\mathbf k} \cdot {\mathbf p} _{nn^\prime}}{m_0} A_{n^\prime }\\
+ \sum_{n^{\prime }} \left(
H^{(1so)}_{nn^{\prime }}
+f\left( {\mathbf r}\right)H^{(\delta so)}_{nn^{\prime }}
+f \left( {\mathbf r}\right) \delta U_{nn^{\prime }}
\right)
A_{n^\prime } =\epsilon A_n,
}
\end{eqnarray}
where
\begin{eqnarray}
H^{(1so)}_{nn^{\prime }} = \frac{\hbar^2\left[ \boldsymbol\nabla U_1\times {\mathbf k}\right] _{nn^{\prime }}\cdot \boldsymbol\sigma }{4m_0^2c^2},
\end{eqnarray}
and
\begin{eqnarray}
H^{(\delta so)}_{nn^{\prime }} = \frac{\hbar^2\left[ \boldsymbol\nabla \delta U\times 
{\mathbf k}\right] _{nn^\prime }\cdot \boldsymbol\sigma} {4m_0^2c^2}.
\end{eqnarray}
We define the matrix elements: ${\mathbf p}_{nn'}=\left\langle n \mid \hbar {\mathbf k}\mid n' \right\rangle $, $\delta U_{nn^{\prime }}=\left\langle n\mid \delta U\mid n^{\prime }\right\rangle$ and $\left[ {\boldsymbol\nabla }U_1\times {\mathbf k}\right] _{nn^{\prime }} \equiv \left\langle n\mid \left[ {\boldsymbol\nabla }U_1\times {\mathbf k}\right] \mid n^{\prime }\right\rangle $. The all-band system of equations (\ref{kp}) is valid for slowly varying envelope functions $A_n$ \cite{Bir-Pikus}.

We do not explicitly include the strain Hamiltonian into consideration, important for lattice-mismatched pairs, in particular for GaN/AlN. The proper procedure, detailed, e.g., in \cite{Zhang} (see also \cite{Lassen2}), would lead to redundant complications, not essential for our results. It suffices to take into account here that the piezoelectric field due to strain along with the possible spontaneous polarization field has contributed to the `external' potential $W\left( {\mathbf r} \right)$. For attainable values of strain, the influence of the deformation potentials on the band edge energies $\epsilon _{n0}$, `offsets' $\delta U_{nn^{\prime }}$ and the matrix elements $H^{(1so)}_{nn^{\prime }}$ and $H^{(\delta so)}_{nn^{\prime }}$ is too weak to be included in the spin-orbit terms being derived.

Ignoring the central cell corrections, some of them contributing to the spin-orbit interaction \cite{Takhtamirov}, we should be aware that there is a number of heterointerface-related effects that cannot be accounted for \cite{Zunger}. The corresponding material parameters are generally not expressed via bulk parameters of the constituent materials, and depend on microscopic structure of the heterointerface and its crystalline orientation \cite{Takhtamirov_Nano}. Their estimates are scarce \cite{Ivchenko,Foreman,Klipstein}, and evidence that they produce a noticeable contribution to the spin-orbit interaction is currently absent. Possible speculations that they alone could explain the huge electron spin splitting in GaN/AlN quantum wells \cite{Lisesivdin} have yet to have some grounds in first-principle band-structure calculations.

We also omitted the $k$-linear spin-orbit interaction term due to the potential $U_1$, see \cite{Dresselhaus,Voon}. In the third perturbation order (with two operators $\hbar {\mathbf k} {\mathbf p}_{nn'}/m_0$), it generates $k$-cubic bulk inversion asymmetry spin-orbit interaction terms, which are conventionally called the Dresselhaus term in zinc-blend materials and the Rashba term for wurtzite \cite{Weber}. Their expressions are known \cite{Dresselhaus,Wang}. One could allow for the position dependence of these terms originating from a $k$-linear spin-orbit interaction contribution due to the potential $f(\mathbf r)\delta U$, which is also omitted from equation (\ref{kp}). Its inclusion may be consistent only for special cases of weakly localized electron states, where the barrier penetration is very essential. This term describes bulk inversion asymmetry position-dependent spin-orbit interaction. Also, we will not consider $k$-linear bulk and the related position-dependent terms for wurtzite structures appearing through the second-order perturbation (due to $H^{(1so)}_{nn^{\prime }} +f\left( {\mathbf r}\right)H^{(\delta so)}_{nn^{\prime }}$ and $\hbar {\mathbf k} {\mathbf p}_{nn'}/m_0$) on the same grounds as above: the expression for the bulk term is known, see~\cite{Voon}, and an analogous position-dependent term has a very week effect on strongly localized states.

\section{Single-band spin-orbit Hamiltonian}

To allow for the interface spin-orbit interaction, we should deal with at least the third-order perturbation term that is a `product' of $f\left( {\mathbf r}\right) H^{(\delta so)}_{nn^{\prime }}$ and two $\hbar {\mathbf k} {\mathbf p}_{nn'}/m_0$. For a simple conduction band with the index $m=s$, the correction of the 3rd order in perturbation ${\bf H}^{\prime }$ is, see Appendix A, equation~(\ref{tildeH}):
\begin{equation}
\tilde H^{(3)}_{ss} = \frac 12\left\{ {\mathbf H}_2,{\mathbf S}_2\right\}_{ss},
\label{H3}
\end{equation}
where the braces stand for a commutator, and ${\mathbf H}_2$ is a part of the perturbation $\mathbf H^{\prime}$ having only non-diagonal couplings of the $s$-band with remote bands; ${\mathbf S}_2$ is given by (\ref{S2ex}). Using $H^{(\delta so)}_{ss} = {\mathbf p}_{ss}=0$, which holds both for zinc-blende and wurtzite materials at the $\Gamma$ point of the Brillouin zone, with the help of expression (\ref{S2ex}), we have:
\begin{equation} \label{H3_so}
\tilde H^{(3)}_{ss}= \sum_{l,l'}
\frac {
H^{\prime}_{sl}
H^{\prime}_{ll'}
H^{\prime}_{l's}
}
{\omega_{sl} \ \omega_{sl'}}.
\end{equation}

To obtain the SIA spin-orbit term originating due to the external potential $W({\mathbf r})$ supplemented with the crystalline potential $f \left( {\mathbf r}\right) \delta U_{nn^{\prime }}$, we deal with the fourth perturbation order in a product of $H^{(1so)}_{nn^{\prime }}$, $W({\mathbf r})\delta_{nn^\prime} + f \left( {\mathbf r}\right) \delta U_{nn^{\prime }}$ and two $\hbar {\mathbf k} {\mathbf p}_{nn'}/m_0$. Here $\delta_{nn^\prime}$ is the Kronecker delta. We do not consider a contribution from $f\left( {\mathbf r}\right)H^{(\delta so)}_{nn^{\prime }}$ as put in the place of $H^{(1so)}_{nn^{\prime }}$. It may produce a noticeable effect only if two conditions are satisfied: we treat weakly localized states with high enough probability for the electron to be in the barrier material, and $H^{( \delta so)}_{nn^{\prime }} \sim H^{(1so)}_{nn^{\prime }}$. The latter condition would mean that the interface contribution (\ref{H3_so}) dominated over the electric field-induced one (in terms of section~2, $\delta \Delta \sim \Delta$) canceling the very consideration of anything farther than the third perturbative order.

If we included the spin-orbit interaction $\mathbf H^{(1so)}$ in the zero-order Hamiltonian ${\mathbf H}_0$, we would not have to go down to the fourth perturbative order, as only the third will be required \cite{Leibler-so}, expressed as a product of $W({\mathbf r}) + f \left( {\mathbf r}\right) \delta U_{nn^{\prime }}$ and two $\hbar {\mathbf k} {\mathbf p}_{nn'}/m_0$. Then, to be consistent, we would have to use the spinor basis functions $u^{so}_{n0}$ that could not be treated as zero-order combinations of the functions $u_{n0}$ \cite{Luttinger-Kohn}. Instead, they should be constructed using not less than the first-order functions:
\begin{equation}
\tilde u^{so}_{n0}=u_{n0}+{\sum_{n' }}^{\prime}
\frac{ H^{(1so)}_{n^{\prime }n}}
{\omega_{nn'}}\,u_{n^{\prime }0}, \label{left-right}
\end{equation}
where the summation does not include degenerate terms. Using the functions (\ref{left-right}) with the third-order perturbative expressions \cite{Leibler-so}, the final result will still have a character of the fourth-order smallness. We consistently treat $\mathbf H^{(1so)}$ as a perturbation to conform with the conventional classification of the basis functions.

The correction of the fourth order is, see appendix~A, equation~(\ref{tildeH}),
\begin{equation} \label{H4}
\tilde H^{(4)}_{ss}=\frac 12\left\{ {\mathbf H}_2,{\mathbf S}_3\right\}_{ss} -\frac 1{24}\left\{ \left\{
\left\{ {\mathbf H}_2,{\mathbf S}_1\right\} ,{\mathbf S}_1\right\} ,{\mathbf S}_1\right\}_{ss},
\end{equation}
where ${\mathbf S}_1$ and ${\mathbf S}_3$ are given by (\ref{S1ex}) and (\ref{S3ex}). Again, using the properties $H^{(1so)}_{ss} = {\mathbf p}_{ss}=0$ and the expressions (\ref{S1ex}) and (\ref{S3ex}), we have
\begin{eqnarray}\label{H4_so}
\eqalign{
\tilde H^{(4)}_{ss}= \sum_{l,l',l''}&
\frac {
H^{\prime}_{sl}
H^{\prime}_{ll'}
H^{\prime}_{l'l''}
H^{\prime}_{l''s}
}
{\omega_{sl} \ \omega_{sl'}\ \omega_{sl''}} 
-\frac 12 \sum_{l,l'}
\left(
\frac 1 {\omega^2_{sl} \ \omega_{sl'}}
+\frac 1 {\omega_{sl} \ \omega^2_{sl'}}
\right)\\
&\times \left(
H^{\prime}_{sl}
H^{\prime}_{ll'}
H^{\prime}_{l's}
H^{\prime}_{ss}+
H^{\prime}_{ss}
H^{\prime}_{sl'}
H^{\prime}_{l'l}
H^{\prime}_{ls}+
H^{\prime}_{sl}
H^{\prime}_{ls}
H^{\prime}_{sl'}
H^{\prime}_{l's}
\right). 
}
\end{eqnarray}
Another property we will use below is that the matrix element $\delta U_{nn'}$ is finite only for functions $\vert n \rangle$ and $\vert n' \rangle$ of the same symmetry. 

\subsection{Zinc-blende semiconductors}

For the conduction band in structures based on zinc-blende semiconductors, expressions (\ref{H3_so}) and (\ref{H4_so}) are greatly simplified as we can neglect the matrix elements $H^{(1so)}_{sl}$ and $H^{(\delta so)}_{sl}$ for any $l$. To prove it, we use the following arguments. Firstly, the spin-orbit interaction operator can be written as the product of the spin and orbital angular momentum near the atomic nuclei, where the interaction is essential. Secondly, the zone-centre function $u_{s0}$, which transforms accordingly to the $\Gamma_1$ representation of the space group ${\text T}_d$, is composed of spherically symmetric atomic $s$ orbitals with zero angular momentum.

The functions $u_{n0}$ can be chosen real, so that
\begin{eqnarray}
\delta U_{nn'} = \delta U_{n'n}, \quad {\mathbf p}_{nn'}=-{\mathbf p}_{n'n}, \quad H^{(1so)}_{nn'} = - H^{(1so)}_{n'n}.
\label{real_ident}
\end{eqnarray}
Using these properties and changing the band summation indices, we have the following for the third perturbative order from (\ref{H3_so}): 
\begin{equation}
\tilde H^{(3)}_{ss} = \sum_{l,l'}
\frac
{
\hbar^2 (p_\alpha)_{sl}\,H^{(\delta so)}_{ll'}\, (p_\beta)_{l's}
}
{2im^2_0 \ \omega_{sl} \ \omega_{sl'}}
\left[ \nabla_\alpha f \, k_\beta - \nabla_\beta f\, k_\alpha \right],
\end{equation}
where $\alpha,\beta = x,y,z$, and summation over these indices is implied here and henceforth.

For the fourth perturbative order, only the first term on the right-hand side of equation~(\ref{H4_so}) gives a finite contribution:
\begin{eqnarray}
\eqalign{
\tilde H^{(4)}_{ss}= &\sum_{l,l'}
\frac
{
\hbar^2 (p_\alpha)_{sl}\,H^{(1so)}_{ll'}\, (p_\beta)_{l's}
}
{im^2_0 \ \omega^2_{sl} \ \omega_{sl'}}
\left[ \nabla_\alpha W \, k_\beta - \nabla_\beta W \, k_\alpha \right]\\
&+\sum_{l,l',l''}
\frac
{
\hbar^2 (p_\alpha)_{sl}\,\delta U_{ll'}\,H^{(1so)}_{l'l''}\, (p_\beta)_{l''s}
}
{im^2_0 \ \omega_{sl} \ \omega_{sl'} \ \omega_{sl''}}
\left[ \nabla_\alpha f \, k_\beta - \nabla_\beta f \, k_\alpha \right].
}
\end{eqnarray}

Finally, we arrive at the traditionally looking spin-orbit interaction terms \cite{Landavshitz}, now generated by both the external scalar potential and variation in the chemical composition of the structure, in the form first given in \cite{Leibler-so}, $H_{so} = \tilde H^{(3)}_{ss} + \tilde H^{(4)}_{ss}$:
\begin{eqnarray}\label{zb}
H_{so} = 
R_{1ZB} \left[ \boldsymbol\nabla f \left( \mathbf r \right) \times {\mathbf k} \right] \cdot \boldsymbol\sigma +
R_{2ZB} \left[ \boldsymbol\nabla W \left( \mathbf r \right) \times {\mathbf k} \right] \cdot \boldsymbol\sigma,
\end{eqnarray}
where
\begin{eqnarray}\label{R1ZB}
\eqalign{
R_{1ZB} = &\sum_{l,l'} 
\frac{
\hbar^4 (p_x)_{sl} \,
\left( \left[ {\boldsymbol\nabla }\delta U\times {\mathbf k}\right]_z\right)_{ll'} \,
(p_y)_{l's}}
{4im^4_0\,c^2 \omega_{sl} \ \omega_{sl'}}\\
&+\sum_{l,l',l''} \frac
{ \hbar^4 (p_x)_{sl} \, \delta U_{ll'} \,
\left( \left[ {\boldsymbol\nabla }U_1\times {\mathbf k}\right]_z\right)_{l'l''} \,
(p_y)_{l''s}} {2im^4_0\,c^2 \omega_{sl} \ \omega_{sl'} \ \omega_{sl''}},
}
\end{eqnarray}
and
\begin{equation}\label{R2ZB}
R_{2ZB} = \sum_{l,l'} \frac
{ \hbar^4 (p_x)_{sl} \,
\left( \left[ {\boldsymbol\nabla }U_1\times {\mathbf k}\right]_z\right)_{ll'} \,
(p_y)_{l's}} {4im^4_0\,c^2}
\left(
\frac 1 {\omega^2_{sl} \ \omega_{sl'}}
+\frac 1 {\omega_{sl} \ \omega^2_{sl'}}
\right).
\end{equation}
Note that the heterointerface contribution, which is proportional to the parameter $R_{1ZB}$, originates not only due to the difference in the spin-orbit interaction energies, as given with the first term in the right-hand side of (\ref{R1ZB}), but also due to the finite matrix elements $\delta U_{ll'}$, see the second term of (\ref{R1ZB}). All items are present in \cite{Leibler-so}, but the latter, because of the chosen spinor basis, contributed to the position-dependent effective mass and was not analysed.

For illustrative purposes, let us limit ourselves to the truncated eight-band Kane model, with the degenerate valence band edge's functions $u_{X0}$, $u_{Y0}$ and $u_{Z0}$ transforming as $x$, $y$ and $z$, respectively, in accordance with the $\Gamma_{15}$ representation of the space group ${\text T}_d$. If we introduce $P = i \left\langle s \mid \hbar k_x\mid X \right\rangle$, $E_g = \omega_{sX}$, $\delta U_v = \delta U_{XX}$ and
\begin{eqnarray}
\frac{\Delta_{so}}{3i}=
\frac
{
\hbar^2 \left( \left[ {\boldsymbol\nabla }U_1\times {\mathbf k}\right]_z\right)_{XY}}
{4m^2_0\,c^2},\quad
\frac{\delta \Delta_{so}}{3i}=
\frac
{
\hbar^2 \left( \left[ {\boldsymbol\nabla } \delta U \times {\mathbf k}\right]_z\right)_{XY}}
{4m^2_0\,c^2},
\end{eqnarray}
we obtain
\begin{equation}\label{R1}
R_{1ZB} = -
\frac
{\hbar^2 P^2 }
{3m^2_0\,E^2_g}\left( \delta \Delta_{so} + \frac {2 \delta U_v \Delta_{so}}{E_g} \right),
\end{equation}
and
\begin{equation}\label{R2}
R_{2ZB} = -
\frac
{2\hbar^2 P^2 \Delta_{so}}
{3m^2_0\,E^3_g},
\end{equation}
which coincide with the known result (see \cite{Silva,Gerchikov,Pfeffer} and put the eigenenergy entering the Hamiltonians there $\epsilon = \epsilon_{s0}$), for $\Delta_{so}, \vert \delta \Delta_{so} \vert, \vert \delta U_v \vert \ll E_g$ and 1D external electric field. Using GaAs/AlAs band parameters \cite{Vurgaftman}, see also section~2, we have $R_{1ZB} \approx 4$~eV\AA$^2$ and $R_{2ZB} \approx -6$~\AA$^2$. Note the different signs of these parameters. They partially cancel each other for GaAs/AlAs quantum wells, which is seen if we use the identity (\ref{delta-E}). If the conduction band offset were $\delta U_s = - R_{1ZB}/R_{2ZB} \approx 0.7$~eV, they would cancel each other exactly (actually $\delta U_s \approx 1$~eV). In intentionally asymmetric quantum wires and dots with strong confinement, this cancellation will be mitigated, and more pronounced spin splittings will be attained.

\subsection{Wurtzite semiconductors}

For Brillouin zone-centre conduction band states in structures based on wurtzite semiconductors, the band edge function $u_{s0}$ transforms as belonging in the $\Gamma_1$ representation of the space group ${\text C}_{6v}$. It is formed of atomic $s$ and $p$ orbitals, so that $H^{(1so)}_{sl}$ and $H^{(\delta so)}_{sl}$ are finite \cite{Voon}.

The basis functions $u_{n0}$ can still be chosen real, producing the identities (\ref{real_ident}). Then we have, for the 3rd perturbative order from (\ref{H3_so}), 
\begin{eqnarray}\label{H3WZ}
\eqalign{
\tilde H^{(3)}_{ss} = &\sum_{l,l'}
\frac
{
\hbar^2 (p_\alpha)_{sl}\,H^{(\delta so)}_{ll'}\, (p_\beta)_{l's}
}
{2im^2_0 \ \omega_{sl} \ \omega_{sl'}}
\left[ \nabla_\alpha f  \, k_\beta - \nabla_\beta f \, k_\alpha \right]\\
&+
\sum_{l,l'}
\frac
{
\hbar^2 (p_\alpha)_{sl}\, (p_\beta)_{ll'} \,H^{(\delta so)}_{l's}
}
{i m^2_0 \ \omega_{sl} \ \omega_{sl'}}
\left[
k_\alpha \left( \nabla_\beta f \right)
+ \left( \nabla_\alpha f \right) k_\beta
\right].
}
\end{eqnarray}
We have used the identity
\begin{equation}
k_\alpha k_\beta \, f - f \, k_\alpha k_\beta = -i k_\alpha \left( \nabla_\beta f \right)
- i \left( \nabla_\alpha f \right) k_\beta .
\end{equation}

For the fourth perturbative order, similar to the zinc-blende case, the second term on the right-hand side of equation~(\ref{H4_so}) gives no contribution. We have
\begin{eqnarray}\label{H4_wz}
\eqalign{
\tilde H^{(4)}_{ss} = &
B_{1\alpha \beta}\left[ \nabla_\alpha W\, k_\beta - \nabla_\beta W\, k_\alpha \right]
+ B_{2\alpha \beta}\left[
k_\alpha \left( \nabla_\beta W \right) +
\left( \nabla_\alpha W \right) k_\beta 
\right]\\
&+ C_{1\alpha \beta}\left[ \nabla_\alpha  f \, k_\beta - \nabla_\beta f\, k_\alpha \right]
+ C_{2\alpha \beta}\left[
k_\alpha \left( \nabla_\beta f \right) +
\left( \nabla_\alpha f \right) k_\beta
\right].
}
\end{eqnarray}
where
\begin{equation}
B_{1\alpha \beta}= \sum_{l,l'}
\frac
{
\hbar^2 (p_\alpha)_{sl}\,(p_\beta)_{ll'}\, H^{(1so)}_{l's}
}
{im^2_0 \ \omega^2_{sl} \ \omega_{sl'}}
+\frac
{
\hbar^2 (p_\alpha)_{sl}\,H^{(1so)}_{ll'}\,(p_\beta)_{l's}
}
{im^2_0 \ \omega^2_{sl} \ \omega_{sl'}},
\end{equation}
\begin{equation}
B_{2\alpha \beta}= \sum_{l,l'}
\frac
{
\hbar^2 (p_\alpha)_{sl}\,(p_\beta)_{ll'}\, H^{(1so)}_{l's} 
}
{im^2_0 \ \omega_{sl} \ \omega^2_{sl'}},
\end{equation}
\begin{equation}
C_{1\alpha \beta}= \sum_{l,l',l''}
\frac
{
\hbar^2 (p_\alpha)_{sl}\, \delta U_{ll'}\, H^{(1so)}_{l'l''}\, (p_\beta)_{l''s}
}
{im^2_0 \ \omega_{sl} \ \omega_{sl'} \ \omega_{sl''}}
+\frac
{
\hbar^2 H^{(1so)}_{sl}\, (p_\alpha)_{ll'}\,\delta U_{l'l''}\,(p_\beta)_{l''s}
}
{im^2_0 \ \omega_{sl} \ \omega_{sl'} \ \omega_{sl''}},
\end{equation}
\begin{equation}
C_{2\alpha \beta}= \sum_{l,l',l''}
\frac
{
\hbar^2 (p_\alpha)_{sl}\, (p_\beta)_{ll'} \, \delta U_{l'l''}\, H^{(1so)}_{l''s} 
}
{im^2_0 \ \omega_{sl} \ \omega_{sl'} \ \omega_{sl''}}
+\frac
{
\hbar^2 H^{(1so)}_{sl}\, (p_\alpha)_{ll'}\,(p_\beta)_{l'l''} \, \delta U_{l''s}
}
{im^2_0 \ \omega_{sl} \ \omega_{sl'} \ \omega_{sl''}},
\end{equation}

Finally, letting the wurtzite $c$-axis be along the $z$-direction, we have $H_{so} = \tilde H^{(3)}_{ss} + \tilde H^{(4)}_{ss}$:
\begin{eqnarray}\label{wz}
\eqalign{
H_{so} =&
R_{1WZ} \left(
\left[ \boldsymbol \nabla f \times {\mathbf k} \right]_x \sigma_x
+ \left[ \boldsymbol \nabla f \times {\mathbf k} \right]_y \sigma_y
\right)+R'_{1WZ} \left[ \boldsymbol \nabla f \times {\mathbf k} \right]_z \sigma_z \\
&+\alpha_1 \left[ \left(\nabla_y f \right) k_z +k_y \left( \nabla_z f\right) \right] \sigma_x -\alpha_1 \left[ \left(\nabla_x f \right) k_z +k_x \left( \nabla_z f \right)\right] \sigma_y\\
&+\alpha_2 \left[ \left(\nabla_x f \right) k_y +k_x \left(\nabla_y f \right) \right] \sigma_z\\
&+R_{2WZ} \left(
\left[ \boldsymbol \nabla W \times {\mathbf k} \right]_x \sigma_x
+ \left[ \boldsymbol \nabla W \times {\mathbf k} \right]_y \sigma_y
\right)+R'_{2WZ} \left[ \boldsymbol \nabla W \times {\mathbf k} \right]_z \sigma_z \\
&+\beta_1 \left[ \left(\nabla_y W \right) k_z +k_y \left( \nabla_z W\right) \right] \sigma_x -\beta_1 \left[ \left(\nabla_x W \right) k_z +k_x \left( \nabla_z W \right)\right] \sigma_y\\
&+\beta_2 \left[ \left(\nabla_x W \right) k_y +k_x \left(\nabla_y W \right) \right] \sigma_z.
}
\end{eqnarray}
The material parameters entering here are given in appendix B. The structure of this complicated expression resembles the net SIA spin-orbit Hamiltonian for zinc-blende systems (\ref{zb}). The differences are plain.  Due to the anisotropy of the wurtzite, $R_{1WZ} \ne R'_{1WZ}$ and $R_{2WZ} \ne R'_{2WZ}$ in the Rashba terms. Another difference consists in the presence of new contributions due to the finite matrix elements of the spin-orbit interaction $H^{(1so)}_{sl}$ and $H^{(\delta so)}_{sl}$ between the conduction and remote bands. New terms are proportional to $\alpha_1$, $\alpha_2$, $\beta_1$ and $\beta_2$, with $\alpha_1 \ne \alpha_2$, $\beta_1 \ne \beta_2$ due to the anisotropy of the wurtzite. These terms are reduced to the conventional form of the Rashba spin-orbit interaction for 1D external electric field and variation in the chemical composition, if we put $\nabla_x W = \nabla_y W = \nabla_x f = \nabla_y f= 0$, $\nabla_z W \ne 0$ and $\nabla_z f  \ne 0$, which is easily seen from expression (\ref{wz}).

It is interesting to learn which bands contribute to make $\alpha_1$, $\alpha_2$, $\beta_1$ and $\beta_2$ finite (see appendix B). It can be deduced using the tables of direct products of irreducible representations of the space group ${\text C}_{6v}$, see \cite{Heine}: $\Gamma_6 \times \Gamma_6 = \Gamma_1 + \Gamma_2 + \Gamma_6$, $\Gamma_2 \times \Gamma_6 = \Gamma_6$, and $\Gamma_1 \times \Gamma_j = \Gamma_j$ for any $j$. We should also take into account that polar vectors (e.g. $\mathbf k$) transform as $\Gamma_1 + \Gamma_6$, while axial vectors (e.g. $\left[ \boldsymbol\nabla U_1 \times \mathbf k\right]$ ) transform as $\Gamma_2 + \Gamma_6$ \cite{Bir-Pikus}. Then we immediately conclude that bands with symmetries $\Gamma_1$ and $\Gamma_6$ define the strength of the parameters $\alpha_1$ and $\beta_1$. Hence, they are finite even in the truncated bands Kane-like model with the nearest valence bands $\Gamma^v_6$ and $\Gamma^v_1$. The term involves new matrix elements, not expressed via known band parameters \cite{Vurgaftman}: $\left\langle \Gamma^v_6\mid \hbar k_y\mid \Gamma^v_1 \right\rangle $ and $\left\langle s\mid \left[ {\boldsymbol\nabla }U_1\times {\mathbf k}\right]_x \mid \Gamma^v_6 \right\rangle $, see also \cite{Voon}.

For parameters $\alpha_2$ and $\beta_2$, in a similar way, there should be finite contributions from the states with symmetries $\Gamma_2$ and $\Gamma_6$. This turns $\alpha_2$ and $\beta_2$ to zero in the truncated bands model. The bands $\Gamma_2$ do not appear in the pseudopotential calculations \cite{Rubio,Beresford}, which probably means that they are very remote.

Let us consider only the nearest valence bands $\Gamma^v_6$ and $\Gamma^v_1$ to estimate $R_{1WZ}$, $R'_{1WZ}$, $R_{2WZ}$ and $R'_{2WZ}$ for GaN/AlN. We can use the expressions (\ref{R1}) and (\ref{R2}). As previously,  the material parameters are taken from \cite{Vurgaftman}. We obtain $R_{2WZ} = R'_{2WZ} \approx -0.01$~\AA$^2$. Accidentally, due to different signs and comparable strength of the terms entering the parenthesis of expression (\ref{R1}), we have vanishingly small $R_{1WZ} = R'_{1WZ} \approx 0.7$~meV\AA$^2$. The external electric field-induced spin orbit interaction dominates because the conduction band offset $\delta U_s \approx 2$~eV, so that $\vert R_{2WZ} \delta U_s \vert = 20$~meV\AA$^2 \gg R_{1WZ}$. While the value of the effective Rashba parameter for GaN/AlN is only one-hundredth of that for the GaAS/AlAs system, very strong electric fields acting on electrons in GaN/AlN can induce large spin splitting of electron states, comparable to that in narrow-bandgap materials \cite{Litvinov}.  

\section{Conclusions}

In the $\mathbf {k \cdot p}$ method, we derived SIA spin-orbit interaction terms for conduction band states near the Brillouin zone centre in zinc-blende and wurtzite semiconductor heterostructures taking into account all remote bands. The results are applicable to quantum wells, wires or dots. Electric field-induced terms and heterointerface contributions were considered, both generally having comparable strength. They can be written in a unified manner only for 2D systems. The resulting expression for the spin-orbit Hamiltonian (\ref{zb}) in zinc-blende materials takes the conventional form of the relativistic spin-orbit interaction \cite{Landavshitz}. The Kane model is adequate to establish this form, with other remote bands only changing the values of the material parameters. For wurtzite materials, the net SIA spin-orbit Hamiltonian (\ref{wz}) has a complicated form due to the anisotropy of the wurtzite and new contributions, which appear owing to finite matrix elements of the spin-orbit interaction $H^{(1so)}_{sl}$ and $H^{(\delta so)}_{sl}$ between the conduction and remote bands. Knowledge of the parameters of the Kane model alone is insufficient to write the spin-orbit Hamiltonian for wurtzite. The effect of remote bands is yet to be evaluated. 

We analysed two popular semiconductor pairs, GaAs/AlAs and GaN/AlN, with the goal of establishing the mechanisms actually governing SIA spin-orbit interaction in heterostructures composed of these materials. Both pairs have `accidental' sets of parameters strongly differing from the `general' picture. For 2D systems GaAs/AlAs, the interface and external electric field-induced contributions are comparable and have different signs partially canceling each other and significantly reducing the net spin splitting. In asymmetric quantum wires and dots with strong enough confinement, this cancellation will be mitigated, and relatively more pronounced spin splittings can be attained. For the systems based on GaN/AlN, the nominally interface-induced contribution is very small as compared to the external electric field-induced one. The evaluations were based on parameters available for the Kane model alone \cite{Vurgaftman}, without remote bands.

\section*{Acknowledgement}

We are grateful to the referees for their comments and suggestions, which stimulated considerable improvement of the content. This work was supported by the NSERC and CRC Program, Canada.

\appendix
\section{L\"owdin perturbation scheme up to the fourth order}

We sketch the L\"owdin perturbation scheme \cite{Lowdin}, which is an efficient tool for the $\mathbf{k\cdot p}$ diagonalization treatment \cite{Leibler-so,Leibler,Lassen,Foreman2}. We follow \cite{Bir-Pikus} and then derive all necessary elements for the fourth-order correction.
The ${\bf k\cdot p}$ system (\ref{kp}) can be presented as:
\begin{equation}
\left( {\mathbf H}_0+{\mathbf H}^{\prime }\right) {\mathbf A}=\epsilon {\mathbf A.}
\end{equation}
Here ${\mathbf H}_0$ is the Hamiltonian of the zero-order approach:
\begin{equation}
H_{0nn^{\prime }} =\epsilon _{n0}\delta _{nn^{\prime }},
\end{equation}
where $\delta _{nn^{\prime }}$ is the Kronecker delta, and ${\mathbf H}^{\prime }={\mathbf H}_1+{\mathbf H}_2$ is the perturbation. The all-band Hamiltonian is being decomposed into the block of $m$-indexed states, whose mutual interaction is taken into account exactly, and a block of `remote' $l$-indexed bands, treated as a perturbation. Here our $m$-class block consists of only a single conduction band $m=s$, but we are preserving the general notations with indices $m$, $m'$, $m''$, etc all belonging to the $m$ class, following \cite{Bir-Pikus}. The perturbation ${\mathbf H}_1$ does not contain elements of interaction of $m$-indexed bands with other bands. In other words, it is already block-diagonal. It is these elements of $m$-$l$ band interaction that are contained in ${\mathbf H}_2$.

The canonical transformation of the envelope functions
\begin{equation}
{\bf \tilde A}=e^{-{\bf S}}{\bf A} 
\end{equation}
with an anti-Hermitian (${\mathbf S}^+ = - {\mathbf S}$) matrix ${\mathbf S}$ results in a set of equations
\begin{equation}
\mathbf {\tilde H\ \tilde A}=\epsilon \mathbf {\tilde A}
\end{equation}
with
\begin{equation}
{\bf \tilde H=}e^{-{\bf S}}{\bf H}e^{{\bf S}}. 
\end{equation}
The properly chosen ${\mathbf S}$ must provide a decomposition of the whole system into a set of equations for $m$-bands and an abandoned set of equations for other bands. Expanding $\exp({\mathbf S})$ in a series
\begin{eqnarray}
e^{\mathbf S}=1+{\mathbf S}+\frac 12{\mathbf S}^2+\frac 1{3!}{\mathbf S}^3+..., 
\end{eqnarray}
we obtain
\begin{eqnarray}
\tilde {\mathbf H}=\sum\limits_{n=0}^\infty \frac 1{n!}\left\{ \mathbf H,\mathbf S\right\}
^{\left( n\right) }. 
\end{eqnarray}
Here
\begin{eqnarray}
\left\{ \mathbf H,\mathbf  S \right\} ^{\left( 0\right) }=\mathbf H, \quad \left\{ \mathbf 
{H,S}\right\} ^{\left( 1\right) }=\left\{ \mathbf {H,S}\right\}, 
\end{eqnarray}
\begin{eqnarray}
\left\{ {\bf H,S}\right\} ^{\left( 2\right) }=\left\{ \left\{ {\bf H,S}%
\right\} ,{\bf S}\right\} ,...
\end{eqnarray}
Let us consider
\begin{eqnarray}
{\bf S=S}_1+{\bf S}_2+{\bf S}_3, 
\end{eqnarray}
where ${\mathbf S}_n$ is the matrix of the $n$th order in ${\bf H}^{\prime }$. The elements ${\mathbf S}_n$ are defined using the recursive equations
\begin{eqnarray}
\left\{ {\mathbf H}_0,{\mathbf S}_1\right\} + {\mathbf H}_2 =0, \label{S1}
\end{eqnarray}
\begin{eqnarray}
\left\{ {\mathbf H}_0,{\mathbf S}_2\right\} + \left\{ {\mathbf H}_1,{\mathbf S}_1\right\} =0, \label{S2}
\end{eqnarray}
\begin{eqnarray}
\left\{ {\mathbf H}_0,{\mathbf S}_3\right\} + \left\{ {\mathbf H}_1,{\mathbf S}_2\right\}
+ \frac 13\left\{ \left\{ {\mathbf H}_2,{\mathbf S}_1\right\} ,{\mathbf S}_1\right\} =0. \label{S3}
\end{eqnarray}
Then the transformed Hamiltonian $\tilde {\mathbf H}$ takes the form
\begin{eqnarray}
\eqalign{
\tilde {\mathbf H} = &{\mathbf H}_0+{\mathbf H}_1+\frac 12\left\{ {\mathbf H}_2,{\mathbf S}_1\right\}
+\frac 12\left\{ {\mathbf H}_2,{\mathbf S}_2\right\}\\
&+\frac 12\left\{ {\mathbf H}_2,{\mathbf S}_3\right\} -\frac 1{24}\left\{ \left\{
\left\{ {\mathbf H}_2,{\mathbf S}_1\right\} ,{\mathbf S}_1\right\} ,{\mathbf S}_1\right\} 
}
\label{tildeH}
\end{eqnarray}
to the fourth order in ${\mathbf H}^{\prime }$ inclusively, having no elements of $m$-$l$ band interaction.  It is interesting to note that an apparently similar perturbation method given by Luttinger and Kohn \cite{Luttinger-Kohn} actually differs from the L\"owdin's in an attempt to make a redundant diagonalization inside the remote bands block.

The matrices ${\mathbf S}_1$ and ${\mathbf S}_2$, which are derived from equations~(\ref{S1}) and (\ref{S2}), are known \cite{Bir-Pikus}:
\begin{eqnarray}
S_{1ml}=-\frac {H^{\prime}_{ml}}{\omega_{ml}},\label{S1ex}
\end{eqnarray}
where $\omega_{nn'} = \epsilon _{n0}-\epsilon _{n'0}$, and
\begin{eqnarray}
S_{2ml}=\sum_{m^{\prime }} \frac {H^{\prime}_{mm'} H^{\prime}_{m'l}}
{\omega_{ml} \ \omega_{m'l}}
- \sum_{l^{\prime }} \frac {H^{\prime}_{ml'} H^{\prime}_{l'l}}
{\omega_{ml} \ \omega_{ml'}}. \label{S2ex}
\end{eqnarray}

Now using equations~(\ref{S3}), (\ref{S1ex}) and (\ref{S2ex}), along with the anti-Hermiticity of $\mathbf S$, we obtain
\begin{eqnarray}
\eqalign{
S_{3ml}=&
\sum_{m',l'}\frac {H^{\prime}_{ml'} H^{\prime}_{l'm'} H^{\prime}_{m'l}}{3\ \omega_{ml}}
\left( \frac 1 {\omega_{m'l}\ \omega_{m'l'}} + \frac 2 {\omega_{ml'}\ \omega_{m'l}} +\frac 1 {\omega_{ml'}\ \omega_{m'l'}} \right)\\
&+\sum_{m',l'}
\frac {H^{\prime}_{mm'} H^{\prime}_{m'l'} H^{\prime}_{l'l}}
{\omega_{ml}} \left( \frac 1 {\omega_{m'l}\ \omega_{m'l'}} + \frac 1 {\omega_{ml'}\ \omega_{m'l'}} \right)\\
&- \sum_{m',m''}
\frac {H^{\prime}_{mm'} H^{\prime}_{m'm''} H^{\prime}_{m''l}}
{\omega_{ml} \ \omega_{m'l}\ \omega_{m''l}}
- \sum_{l',l''}
\frac {H^{\prime}_{ml'} H^{\prime}_{l'l''} H^{\prime}_{l''l}}
{\omega_{ml} \ \omega_{ml''}\ \omega_{ml'}}.
}\label{S3ex}
\end{eqnarray}

\section{Material parameters entering expression (\ref{wz})}

\begin{eqnarray}\label{R1wz}
\eqalign{
R_{1WZ} = &\sum_{l,l'} 
\frac{
\hbar^4 (p_y)_{sl} \,
\left( \left[ {\boldsymbol\nabla }\delta U\times {\mathbf k}\right]_x\right)_{ll'} \,
(p_z)_{l's}}
{4im^4_0\,c^2 \omega_{sl} \ \omega_{sl'}}\\
&+\sum_{l,l',l''} \frac
{ \hbar^4 (p_y)_{sl} \, \delta U_{ll'} \,
\left( \left[ {\boldsymbol\nabla }U_1\times {\mathbf k}\right]_x\right)_{l'l''} \,
(p_z)_{l''s}} {4im^4_0\,c^2 \omega_{sl} \ \omega_{sl'} \ \omega_{sl''}}\\
&+\sum_{l,l',l''} \frac
{ \hbar^4 (p_y)_{sl} \,
\left( \left[ {\boldsymbol\nabla }U_1\times {\mathbf k}\right]_x\right)_{ll'}
\, \delta U_{l'l''}\,
(p_z)_{l''s}} {4im^4_0\,c^2 \omega_{sl} \ \omega_{sl'} \ \omega_{sl''}}\\
&+\sum_{l,l',l''} \frac
{\hbar^4 \left( \left[ {\boldsymbol\nabla }U_1\times {\mathbf k}\right]_x\right)_{sl} \,
(p_y)_{ll'} \, \delta U_{l'l''}\,
(p_z)_{l''s}} {4im^4_0\,c^2 \omega_{sl} \ \omega_{sl'} \ \omega_{sl''}}\\
&+\sum_{l,l',l''} \frac
{\hbar^4 (p_y)_{sl} \,
\delta U_{ll'}\, (p_z)_{l'l''}\,
\left( \left[ {\boldsymbol\nabla }U_1\times {\mathbf k}\right]_x\right)_{l''s}
} {4im^4_0\,c^2 \omega_{sl} \ \omega_{sl'} \ \omega_{sl''}},
}
\end{eqnarray}
\begin{eqnarray}\label{Rp1wz}
\eqalign{
R'_{1WZ} = &\sum_{l,l'} 
\frac{
\hbar^4 (p_x)_{sl} \,
\left( \left[ {\boldsymbol\nabla }\delta U\times {\mathbf k}\right]_z\right)_{ll'} \,
(p_y)_{l's}}
{4im^4_0\,c^2 \omega_{sl} \ \omega_{sl'}}\\
&+\sum_{l,l',l''} \frac
{ \hbar^4 (p_x)_{sl} \, \delta U_{ll'} \,
\left( \left[ {\boldsymbol\nabla }U_1\times {\mathbf k}\right]_z\right)_{l'l''} \,
(p_y)_{l''s}} {2im^4_0\,c^2 \omega_{sl} \ \omega_{sl'} \ \omega_{sl''}}\\
&+\sum_{l,l',l''} \frac
{\hbar^4 \left( \left[ {\boldsymbol\nabla }U_1\times {\mathbf k}\right]_z\right)_{sl} \,
(p_x)_{ll'} \, \delta U_{l'l''}\,
(p_y)_{l''s}} {2im^4_0\,c^2 \omega_{sl} \ \omega_{sl'} \ \omega_{sl''}},
}
\end{eqnarray}
\begin{equation}\label{a1}
\eqalign{
\alpha _1 = \sum_{l,l'}&
\frac
{
\hbar^4 (p_y)_{sl} \, (p_z)_{ll'} \,
\left( \left[ {\boldsymbol\nabla } \delta U\times {\mathbf k}\right]_x\right)_{l's}
}
{4im^4_0\,c^2\,\omega_{sl} \ \omega_{sl'}}\\
&-
\frac
{
\hbar^4
\left( \left[ {\boldsymbol\nabla } \delta U\times {\mathbf k}\right]_x\right)_{sl}\,
(p_y)_{ll'} \, (p_z)_{l's}
}
{4im^4_0\,c^2\,\omega_{sl} \ \omega_{sl'}
}\\
&+
\frac
{
\hbar^4
\left( \left[ {\boldsymbol\nabla } U_1\times {\mathbf k}\right]_x\right)_{sl}\,
(p_y)_{ll'} \, (p_z)_{l'l''}\, \delta U_{l''s}
}
{4im^4_0\,c^2\,\omega_{sl} \ \omega_{sl'}\ \omega_{sl''} 
}\\
&-
\frac
{
\hbar^4 \delta U_{sl} \, (p_y)_{ll'} \, (p_z)_{l'l''}
\left( \left[ {\boldsymbol\nabla } U_1\times {\mathbf k}\right]_x\right)_{l''s}\,
}
{4im^4_0\,c^2\,\omega_{sl} \ \omega_{sl'}\ \omega_{sl''} 
}\\
&+
\frac
{
\hbar^4 (p_y)_{sl} \, (p_z)_{ll'}\, \delta U_{l'l''}
\left( \left[ {\boldsymbol\nabla } U_1\times {\mathbf k}\right]_x\right)_{l''s}\,
}
{4im^4_0\,c^2\,\omega_{sl} \ \omega_{sl'}\ \omega_{sl''} 
}\\
&-
\frac
{
\hbar^4
\left( \left[ {\boldsymbol\nabla } U_1\times {\mathbf k}\right]_x\right)_{sl}\,
\delta U_{ll'} \, (p_y)_{l'l''} \, (p_z)_{l''s}
}
{4im^4_0\,c^2\,\omega_{sl} \ \omega_{sl'}\ \omega_{sl''} 
},
}
\end{equation}
\begin{equation}\label{a2}
\eqalign{
\alpha _2 = \sum_{l,l'}&
\frac
{
\hbar^4 (p_x)_{sl} \, (p_y)_{ll'} \,
\left( \left[ {\boldsymbol\nabla } \delta U\times {\mathbf k}\right]_z\right)_{l's}
}
{2im^4_0\,c^2\,\omega_{sl} \ \omega_{sl'}}\\
&+
\frac
{
\hbar^4
\left( \left[ {\boldsymbol\nabla } U_1\times {\mathbf k}\right]_z\right)_{sl}\,
(p_x)_{ll'} \, (p_y)_{l'l''}\, \delta U_{l''s}
}
{2im^4_0\,c^2\,\omega_{sl} \ \omega_{sl'}\ \omega_{sl''} 
}\\
&+
\frac
{
\hbar^4 (p_x)_{sl} \, (p_y)_{ll'}\, \delta U_{l'l''}
\left( \left[ {\boldsymbol\nabla } U_1\times {\mathbf k}\right]_z\right)_{l''s}\,
}
{2im^4_0\,c^2\,\omega_{sl} \ \omega_{sl'}\ \omega_{sl''} 
},
}
\end{equation}
\begin{eqnarray}\label{R2wz}
\eqalign{
R_{2WZ} = \sum_{l,l'}&
\frac
{
\hbar^4 (p_y)_{sl} \,
\left( \left[ {\boldsymbol\nabla }U_1\times {\mathbf k}\right]_x\right)_{ll'} \,
(p_z)_{l's}}
{4im^4_0\,c^2}
\left(
\frac 1 {\omega^2_{sl} \ \omega_{sl'}}
+\frac 1 {\omega_{sl} \ \omega^2_{sl'}}
\right)\\
&+
\frac
{
\hbar^4 (p_y)_{sl} \, (p_z)_{ll'} \,
\left( \left[ {\boldsymbol\nabla }U_1\times {\mathbf k}\right]_x\right)_{l's}
}
{4im^4_0\,c^2\,\omega^2_{sl} \ \omega_{sl'}}\\
&+
\frac
{
\hbar^4
\left( \left[ {\boldsymbol\nabla }U_1\times {\mathbf k}\right]_x\right)_{sl}\,
(p_y)_{ll'} \, (p_z)_{l's}
}
{4im^4_0\,c^2\,\omega_{sl} \ \omega^2_{sl'}
},
}
\end{eqnarray}
\begin{eqnarray}\label{Rp2wz}
\eqalign{
R'_{2WZ} = \sum_{l,l'}&
\frac
{
\hbar^4 (p_x)_{sl} \,
\left( \left[ {\boldsymbol\nabla }U_1\times {\mathbf k}\right]_z\right)_{ll'} \,
(p_y)_{l's}}
{4im^4_0\,c^2}
\left(
\frac 1 {\omega^2_{sl} \ \omega_{sl'}}
+\frac 1 {\omega_{sl} \ \omega^2_{sl'}}
\right)\\
&+
\frac
{
\hbar^4 (p_x)_{sl} \, (p_y)_{ll'} \,
\left( \left[ {\boldsymbol\nabla }U_1\times {\mathbf k}\right]_z\right)_{l's}
}
{4im^4_0\,c^2\,\omega^2_{sl} \ \omega_{sl'}}\\
&+
\frac
{
\hbar^4
\left( \left[ {\boldsymbol\nabla }U_1\times {\mathbf k}\right]_z\right)_{sl}\,
(p_x)_{ll'} \, (p_y)_{l's}
}
{4im^4_0\,c^2\,\omega_{sl} \ \omega^2_{sl'}
}.
}
\end{eqnarray}
\begin{equation}\label{b1}
\eqalign{
\beta _1 = \sum_{l,l'}&
\frac
{
\hbar^4 (p_y)_{sl} \, (p_z)_{ll'} \,
\left( \left[ {\boldsymbol\nabla }U_1\times {\mathbf k}\right]_x\right)_{l's}
}
{4im^4_0\,c^2\,\omega_{sl} \ \omega^2_{sl'}}\\
&-
\frac
{
\hbar^4
\left( \left[ {\boldsymbol\nabla }U_1\times {\mathbf k}\right]_x\right)_{sl}\,
(p_y)_{ll'} \, (p_z)_{l's}
}
{4im^4_0\,c^2\,\omega^2_{sl} \ \omega_{sl'}
},
}
\end{equation}
\begin{equation}\label{b2}
\eqalign{
\beta _2 = \sum_{l,l'}&
\frac
{
\hbar^4 (p_x)_{sl} \, (p_y)_{ll'} \,
\left( \left[ {\boldsymbol\nabla }U_1\times {\mathbf k}\right]_z\right)_{l's}
}
{4im^4_0\,c^2\,\omega_{sl} \ \omega^2_{sl'}}\\
&-
\frac
{
\hbar^4
\left( \left[ {\boldsymbol\nabla }U_1\times {\mathbf k}\right]_z\right)_{sl}\,
(p_x)_{ll'} \, (p_y)_{l's}
}
{4im^4_0\,c^2\,\omega^2_{sl} \ \omega_{sl'}
}.
}
\end{equation}


\section*{References}

\begin{thebibliography}{99}

\bibitem{Landavshitz}  V.B.~Berestetskii, E.M.~Lifshitz, and L.P.~Pitaevskii, {\em Quantum Electrodynamics}, 2nd ed. (Perga\-mon Press, Oxford, 1982)

\bibitem{Dresselhaus} G.~Dresselhaus, Phys.\ Rev. {\bf 100}, 580 (1955).

\bibitem{Casella} R.C.~Casella, Phys.\ Rev. {\bf 114}, 1514 (1959).

\bibitem{Rashba1} E.I.~Rashba and V.I.~Sheka, Fiz.\ Tverd.\ Tela (Leningrad), Collection of Articles 2, 162 (1959).

\bibitem{Rashba2} E.I.~Rashba and V.I.~Sheka, in {\em Landau Level Spectroscopy}, G.~Landwehr and E.I.~Rashba, eds. (North-Holland, Amsterdam, 1991).

\bibitem{Bychkov} Yu.A.~Bychkov and E.I.~Rashba, JETP Lett. {\bf 39} 78 (1984).

\bibitem{Silva} E.A.~de~Andrada~e~Silva, G.C.~La~Rocca, and F.~Bassani, Phys.\ Rev.\ B {\bf 55}, 16293 (1997).

\bibitem{Luttinger} J.M.~Luttinger, Phys.\ Rev. {\bf 102}, 1030 (1956).

\bibitem{Bir-Pikus} G.L.~Bir and G.E.~Pikus, {\em Symmetry and Strain-Induced Effects in Semiconductors} (Wiley, New York, 1974).

\bibitem{Dandrea} R.G.~Dandrea, C.B.~Duke, A.~Zunger, J.\ Vac.\ Sci.\ Technol.\ B {\bf 10}, 1744 (1992).

\bibitem{Bernardini} F.~Bernardini and V.~Fiorentini, Phys.\ Rev.\ B {\bf 57}, R9427 (1998).

\bibitem{Zunger} A.~Zunger, phys.\ stat.\ sol.\ (a) {\bf 190}, 467 (2002).

\bibitem{Takhtamirov} E.E.~Takhtamirov and V.A.~Volkov, JETP {\bf  89}, 1000 (1999).

\bibitem{Leibler-so} L.~Leibler, Phys.\ Rev.\ B {\bf 16}, 863 (1977).

\bibitem{Vasko} F.T.~Vas'ko, JETP Lett. {\bf 30}, 541 (1979).

\bibitem{Gerchikov} L.G.~Gerchikov and A.V.~Subashiev, Sov.\ Phys.\ Semicond. {\bf 26}, 73 (1992).

\bibitem{Pfeffer} P.~Pfeffer and W.~Zawadzki, Phys.\ Rev.\ B {\bf 59}, R5312 (1999).

\bibitem{Loss} V.~Cerletti, W.A.~Coish, O.~Gywat, and D.~Loss, Nanotechnology {\bf 16}, R27 (2005).

\bibitem{Gurung} T.~Gurung, S.~Mackowski, G.~Karczewski, H.E.~Jackson, and L.M.~Smith, Appl.\ Phys.\ Lett. {\bf 93}, 153114 (2008).

\bibitem{Benjamin} S.C.~Benjamin, B.W.~Lovett, J.M.~Smith, Laser \& Photon. Rev. {\bf 3}, 556 (2009).

\bibitem{Sanjay} S.~Prabhakar and J.E.~Raynolds, Phys.\ Rev.\ B {\bf 79}, 195307 (2009).

\bibitem{Moroz} A.V.~Moroz and C.H.W.~Barnes, Phys.\ Rev.\ B {\bf 61}, R2464 (2000).

\bibitem{Andreev} A.D.~Andreev and E.P.~O'Reilly, Phys.\ Rev.\ B {\bf 62}, 15851 (2000).

\bibitem{Pan} E.~Pan and B.~Yang, J.\ Appl.\ Phys. {\bf 93}, 2435 (2003).

\bibitem{Barettin} D.~Barettin, B.~Lassen, and M.~Willatzen, J.\ Phys.: Conf.\ Ser. {\bf 107}, 012001 (2008).

\bibitem{Patil1} S.~Patil and R.V.N.~Melnik, Nanotechnology {\bf 20}, 125402, (2009).

\bibitem{Patil2} S.~Patil and R.V.N.~Melnik, Procedia Engineering {\bf 1}, 105 (2009).

\bibitem{Litvinov} V.I.~Litvinov, Phys.\ Rev.\ B {\bf 68}, 155314 (2003).

\bibitem{Lisesivdin} S.B. Lisesivdin, N.~Balkan, O.~Makarovsky, A.~ Patan\`e, A.~Yildiz, M.D.~Caliskan, M.~Kasap, S.~Ozcelik, and E.~Ozbay, J.\ Appl.\ Phys. {\bf 105}, 093701 (2009).

\bibitem{Luttinger-Kohn} J.M.~ Luttinger and W.~Kohn, Phys.\ Rev. {\bf 97}, 869 (1955).

\bibitem{Kane} E.O.~Kane, J.\ Phys.\ Chem.\ Solids, {\bf 1} 249 (1957).

\bibitem{Suris} R.A.~Suris, Sov.\ Phys.\ Semicond. {\bf 20} 1258 (1986).

\bibitem{Lowdin} P.-O.~L\"owdin, J.\ Chem.\ Phys. {\bf 19}, 1396 (1951).

\bibitem{Vurgaftman} I.~Vurgaftman, J.R.~Meyer, and L.R.~Ram-Mohan,  J.\ Appl.\ Phys. {\bf 89}, 5815 (2001).

\bibitem{Leibler}  L.~Leibler, Phys.\ Rev.\ B {\bf 12}, 4443 (1975).

\bibitem{Zhang} Y.~Zhang, Phys.\ Rev.\ B {\bf 49}, 14352 (1994).

\bibitem{Lassen2} B.~Lassen, M.~Willatzen, and R.~Melnik, J.\ Comput.\ Theor.\ Nanosci. {\bf 3}, 588 (2006).

\bibitem{Takhtamirov_Nano} E.E.~Takhtamirov and V.A.~Volkov in {\em Proceedings of 7th International Symposium ``Nanostructures: Physics and Technology''}, St.~Petersburg: Ioffe Institute, 1999, p. 303.

\bibitem{Ivchenko} E.L.~Ivchenko, A.Yu.~Kaminski, and U.M.~R\"ossler, Phys.\ Rev.\ B {\bf 54}, 5852 (1996).

\bibitem{Foreman} B.A.~Foreman, Phys.\ Rev.\ B {\bf 76}, 045327 (2007).

\bibitem{Klipstein} P.C.~Klipstein, Phys.\ Rev.\ B {\bf 81}, 235314 (2010).

\bibitem{Voon} L.C.~Lew~Yan~Voon, M.~Willatzen, M.~Cardona, and N.E.~Christensen, Phys.\ Rev.\ B {\bf 53}, 10703 (1996).

\bibitem{Weber} W.~Weber, S.D.~Ganichev, S.N.~Danilov, D.~Weiss, W.~Prettl,
Z.D.~Kvon, V.V.~Bel'kov, L.E.~Golub, H.-I.~Cho, and J.-H.~Lee, Appl.\ Phys.\ Lett. {\bf 87}, 262106 (2005).

\bibitem{Wang} W.-T.~Wang, C.L.~Wu, S.F.~Tsay, M.H.~Gau, I.~Lo, H.F.~Kao, D.J.~Jang, J.-C.~Chiang, M.-E.~Lee, Y.-C.~Chang, C.-N.~Chen, and H.C.~Hsueh, Appl.\ Phys.\ Lett. {\bf 91}, 082110 (2007).

\bibitem{Lassen} B.~Lassen, R.V.N.~Melnik, and M.~Willatzen, Commun. Comput. Phys., {\bf 6}, 699 (2009).

\bibitem{Foreman2} B.A.~Foreman, Phys.\ Rev.\ Lett. {\bf 84}, 2505 (2000).

\bibitem{Heine} V.~Heine, {\em Group Theory in Quantum Mechanics} (Pergamon Press, New York, 1960).

\bibitem{Rubio} A.~Rubio, J.L.~Corkill, M.L.~Cohen, E.L.~Shirley, and S.G.~Louie, Phys.\ Rev.\ B {\bf 48}, 11810 (1993).

\bibitem{Beresford} R.~Beresford, J.\ Appl.\ Phys. {\bf 95}, 6216 (2004).

\end{thebibliography}
\end{document}